\documentclass[twocolumn,amsmath,amssymb]{revtex4-1}
\usepackage{bm,amsfonts, mathtools}
\usepackage{graphicx,color, wasysym}
\usepackage{textcomp}
\usepackage{amsmath,amssymb,latexsym,epsfig}
\usepackage[british]{babel}
\usepackage{hyphenat}
\usepackage{appendix}

\DeclareMathOperator{\okr}{{\stackrel{{\scriptscriptstyle{\mathsf{def}}}}{=}}}

\DeclareMathOperator{\D}{d\!}
\DeclareMathOperator{\E}{e} 
\DeclareMathOperator{\I}{i}

\begin{document}

\title{Non-Debye relaxations: smeared time evolution, memory effects, \\and the Laplace exponents}

\author{K.~G\'{o}rska}
\email{katarzyna.gorska@ifj.edu.pl}
\author{A. Horzela}
\email{andrzej.horzela@ifj.edu.pl}
\affiliation{H. Niewodnicza\'{n}ski Institute of Nuclear Physics, Polish Academy of Sciences, \\ 
ul.Eljasza-Radzikowskiego 152, PL 31342 Krak\'{o}w, Poland}

\author{T.~K.~Pog\'any}
\email{poganj@pfri.hr}
\affiliation{Faculty of Maritime Studies, University of Rijeka, 51000 Rijeka, Croatia \\
Institute of Applied Mathematics, \'Obuda University, H-1034 Budapest, Hungary}

\begin{abstract}
The non-Debye, \textit{i.e.,} non-exponential, behavior characterizes a large plethora of dielectric relaxation 
phenomena. Attempts to find their theoretical explanation are dominated  either by considerations rooted in the stochastic processes methodology or by the so-called \textsl{fractional dynamics} based on equations involving fractional derivatives which mimic the non-local time evolution and as such may be interpreted as describing memory effects. Using the recent results coming from the stochastic approach  we link memory functions with the Laplace (characteristic) exponents of infinitely divisible probability distributions and show how to relate 
the latter with experimentally measurable spectral functions characterizing relaxation in 
the frequency domain. This enables us to incorporate phenomenological knowledge into the evolution laws. 
To illustrate our approach we consider the standard Havriliak-Negami and Jurlewicz-Weron-Stanislavsky models 
for which we derive well-defined evolution equations. Merging stochastic and fractional dynamics approaches sheds also new light on the analysis of relaxation phenomena which description needs going beyond using the single evolution pattern. We determine sufficient conditions under which such description is consistent with general requirements of our approach.     
\end{abstract}



%

\maketitle

\section{Introduction}\label{sec1}

The Debye relaxation (abbreviated by a subscript $D$) describes the response of an ideal non--interacting 
population of dipoles subjected to an alternating external electric field. The simplest example of such a phenomenon is the step--like outer field which switching off results in the time decay of induced polarization described by the relaxation function $n(t)$. Its time behaviour is governed by the differential equation $\dot{n}_{D}(t) = - \tau^{-1}\, n_{D}(t)$ where $\dot{n}_{D}(t) = \D n_{D}(t)/\D t$ and $\tau$ denotes a material constant called the relaxation time. 
Solution to this elementary equation required to satisfy the initial condition $n(0)=1$ reads $n_{D}(t) = 
\exp(-t/\tau)$, \textit{i.e.}, the usual exponential law. If the system response is not ideal and mutual interaction of dipoles influences the relaxation process then its description based on the Debye pattern does not work any longer 
and to get agreement with experimental data one has to give up the paradigm of exponential decay. Among efforts focused on generalizations of the time evolution equation the overriding example is to replace the 
constant relaxation time $\tau^{-1}$ by a time dependent transition rate function $r(t, \tau)$ \cite{Mainardi18}. This leads to the equation
   \begin{equation}\label{28/05-1}
      \dot{n}(t) = - r(t, \tau)\, n(t) 
   \end{equation}   
whose solution obeying the initial condition $n(0) = 1$ reads $n(t) = \exp\big(-\!\int_{0}^{t} r(\xi, \tau) \D\xi\big)$. 
The transition rate $r(t, \tau) = -\dot{n}(t)/n(t)$ remains constant only for the Debye relaxation and it is a serious challenge how to determine it for other processes. The basic experimental source of information is the broadband dielectric spectroscopy \cite{KremerSchoenhals}. It provides us with phenomenological knowledge on dispersive and absorptive properties of the medium in the frequency range of several decades of  magnitude, next encoded in the complex valued,  frequency dependent, dielectric permittivity $\hat{\varepsilon}(\I\!\omega)$ \cite{Boettcher}. However, if one utilizes these data to find the time dependence of relaxation functions (it may be calculated explicitly using analytically known expressions which fit the data \cite{Gloeckle, RHilfer02a, RGarrappa16, KGorska18}) then immediately encounters the singularity of $r(t, \tau)$ at $t = 0$. The widely used Kohlrausch-Williams-Watts (KWW) model of the stretched exponential time decay $n_{K\!W\!W}(t) = \exp\big(-(t/\tau)^{\alpha}\big)$ leads to $r_{K\!W\!W}(t, \tau) = \alpha\tau (t/\tau)^{\alpha - 1}$ which for physically typical values $\alpha\in(0, 1)$ is obviously singular at $t=0$. The presence of such a singularity means that $r(t, \tau)$ can not be experimentally 
handled, or credibly extrapolated, for short times. Measurements in a neighbourhood close to $t=0$ 
become uncontrolled and unfaithful as may give completely different results for $n_{K\!W\!W}(t)$ even if repeated in the same experimental conditions. Similar situation is met for another 
phenomenological models of non-Debye relaxations within which calculations of $r(t, \tau)$ usually lead to singular ratios of  
functions belonging to the Mittag-Leffler family. Thus, to get the time evolution equations reliable for 
non-Debye relaxation one does need another approach - an alternative, still belonging to the field of macroscopic physics, is to go beyond investigations of the time evolution in terms of simple ordinary differential equations. Among popular and widely accepted proposals  to solve the problem one finds methods of fractional dynamics based on using equations which involve fractional derivatives. This provides us with many valuable results but has also a serious disadvantage: rules and/or laws of fractional dynamics leaves open the question 
which type of fractional derivative should be chosen as the most suitable for the problem under 
consideration and how to relate it to phenomenological observations. The choice is frequently done 
{\it ad hoc} and many a time justified only by the statement that the proposed model is solvable. This is neither 
physically nor mathematically satisfactory and has motivated us to develop a novel scheme which 
starting point is the integral version of Eq. \eqref{28/05-1}
   \begin{equation*}
      n(t) = 1 - \int_0^t r(\xi, \tau)\, n(\xi)\, \D\xi,
   \end{equation*}
made to be non-local in time by introducing the time smearing $r(\xi, \tau)\rightarrow r(t-\xi, \tau)$ 
   \begin{equation}\label{28/05-1b}
      n(t) = 1 - \int_0^t r(t-\xi, \tau)\,n(\xi)\, \D\xi
   \end{equation}
which reflects memory effects modifying the time evolution law. We shall show that merging this with 
the results obtained in the framework of the stochastic processes based approach to relaxation phenomena (for comprehensive literature see {\cite{AStanislavsky15, AStanislavsky16, AStanislavsky17, AStanislavsky19, AStanislavsky19a}}) we are able to built into 
proposed scheme the phenomenological knowledge of relaxation functions coming from the broadband 
dielectric spectroscopy.   

Physicists who investigate relaxation processes using extensively the probabilistic and statistical methods customary are used to treat Eq. \eqref{28/05-1b} as a kinetic equation related to some stochastic process governed by an 
infinitely divisible distribution and to write it down in the form of the master equation
   \begin{equation}\label{15/04-1}
      n(t)= 1 - B(\tau, {\bf p})\int_{0}^{t} M(t - \xi)\, n(\xi)\, \D\xi,
   \end{equation}
separating out the memory function $M(t)$ and the transition rate constant {$B(\tau, 
{\bf p}) \okr B$ which, besides of the relaxation time, may involve also other parameters  
${\bf p}$}. In the Laplace domain the solution of Eq. \eqref{15/04-1} is  
   \begin{equation*}
      \hat{n}(s)= \{s\,[1 + B\hat{M}(s)]\}^{-1}.
   \end{equation*}
Thus it is the knowledge of the Laplace transformed memory function $\hat{M}(s)$ being supposed to play a role of quantity   
essential for further considerations. Indeed, it does play - the formalism developed within the 
stochastic approach to relaxation processes allows one to connect $\hat{M}(s)$ with the Laplace exponent
 $\hat{\Psi}(s)$ \cite{footnote1} determined by the probability 
distribution governing the stochastic process \cite{AStanislavsky15, AStanislavsky19}). The expected relation 
takes the form
   \begin{equation}\label{29/05-2}
      \hat{M}(s) = [\hat{\Psi}(s)]^{-1},
   \end{equation}
and links dynamical and stochastic aspects of the relaxation phenomena according to
\begin{equation}\label{17/11-1}
       \hat{n}(s)= \{s\,[1 + B/\hat{\Psi}(s)]\}^{-1}.
   \end{equation}
The relations Eq. \eqref{29/05-2} and consequently Eq. \eqref{17/11-1} are unique and open the way to 
use methods worked out in the theory of positive definite functions (\cite{Berg,RSchilling,RSchilling10}, 
for a brief introduction to their properties see {Appendix A}). From the L\'{e}vy--Khintchine 
formula \cite{Berg,RSchilling,RSchilling10} we learn that for $s \in \mathbb  R_>$, $\hat{\Psi}(s)$ is 
a complete Bernstein function (CBF). Properties of CBFs imply also that $s/\hat{\Psi}(s) = \hat{\Phi}(s)$ 
is a CBF as well and that there exists a completely monotone function (CMF) associated with CBF - this 
comes out directly from the fact that the algebraic reciprocal of any CBF is CMF  \cite{footnote2,
footnote2a}. Collecting these facts together we conclude that when describe the relaxation we deal with two couples of functions: the first of them, related to Eq. \eqref{15/04-1}, consists of the memory function $M(t)$ and the Laplace exponent $\hat{\Psi}(s)$, whereas the second one 
involves the Laplace exponent $\hat{\Phi}(s)$ and a new memory function ${k}(t)$ which, through the Laplace transform, corresponds to $\hat{\Phi}(s)$ according to
   \begin{equation*}
      \hat{k}(s) = [\hat{\Phi}(s)]^{-1} = \hat{\Psi}(s)/s.
   \end{equation*}
It is natural to ask the question - what is the equation related to the function $k$? From the just quoted properties of CBFs we know that for $s$ restricted to $\mathbb R_>$,  $\hat{k}(s)$ is CMF. So its time domain analogue $k(t)$ may be interpreted as the memory function which defines the well-posted Cauchy problem 
   \begin{equation}\label{29/05-3}
      \int_{0}^{t} k(t-\xi)\, \dot{\nu}(\xi)\, \D\xi = -B \nu(t)
   \end{equation}
with the initial condition $\nu(0) = 1$ \cite{ANKochubei11}. Eqs. \eqref{15/04-1} and \eqref{29/05-3} stem from the same stochastic process and as such should be read out as mutually alternative descriptions of the same physical reality. Thus we can identify the relaxation functions $n(t)$ and $\nu(t)$. To trace this step better note that the Laplace exponents $\hat{\Psi}(s)$ and $\hat{\Phi}(s)$ are joined by $\hat{\Psi}(s)\,\hat{\Phi}(s) = s$ \cite{AStanislavsky20}. Rewriting this relation for the memory 
kernels in the Laplace domain we obtain
   \begin{equation}\label{30/05-2}
      \hat{k}(s) \hat{M}(s) = s^{-1},
   \end{equation}
which, by the usual convolution property, in the time domain equals to
   \begin{equation}\label{22/07-1}
      \int_{0}^{t} k(\xi)\, M(t-\xi)\, \D\xi = \int_{0}^{t} k(t-\xi)\, M(\xi)\, \D\xi = 1
   \end{equation}
\textit{i.e.}, gives the condition which says that the memory kernels $k(t)$ and $M(t)$ constitute the Sonine pair 
\cite{AHanyga20, KGorska20}, the property long--time noticed and applied in the theory of integral equations. Comparing Eqs. \eqref{15/04-1} and/or \eqref{29/05-3} with Eq. \eqref{28/05-1} we see that they describe the time smearing $r(t, \tau) n(t)$ and/or $\dot{n}(t)$, respectively. As just remarked from the physical point of view both smearings should give the same results, \textit{i.e.,} starting from Eq. \eqref{15/04-1} we should be able to obtain Eq. \eqref{29/05-3} and \textit{vice versa}. Simultaneously, we should recover essential objects describing the relaxation, namely the memory, relaxation and spectral functions if the Laplace exponent of its underlying stochastic process is known. In the view of that the crucial result comes to play - the Laplace exponent is uniquely connected  to the spectral function which real and imaginary parts are measured in dielectric spectroscopy and provide us with basic empirical information on the relaxation process being studied. 

Forthcoming part of the paper is devoted to 
the explanation how just sketched construction works and how it may be illustrated on standard examples of 
the Cole-Cole (CC), Havriliak-Negami (HN) and Jurlewicz-Weron-Stanislavsky (JWS) relaxation patterns. More precisely,
the article is devoted to deriving explicit forms of Eqs. \eqref{15/04-1} 
and \eqref{29/05-3} adjusted to the HN and JWS models. In Sec. \ref{sec2} we recall some basic facts 
concerning the HN and JWS models and identify relaxation functions relevant for these models. The time domain memory functions are calculated in Sec. \ref{sec3} giving results used in 
Sec. \ref{sec4} to write down explicitly Eqs. \eqref{15/04-1} and \eqref{29/05-3} as well as their solutions. 
In Sec. \ref{sec5} we propose an extension of our scheme to encounter processes which description goes beyond using single relaxation pattern - these we shall call ``multichannel'' processes. We show that our approach remains self-consistent for a process governed by a sum of the Cole-Cole patterns parametrized by $\alpha$ and $\beta$ satisfying $\alpha + \beta = 1$. Such a conclusion is expected in view of results obtained recently within the stochastic approach \cite{AStanislavsky20}. The paper is resumed in Conclusions and completed by Appendices added to clarify some not widely known mathematical aspects of the exposition.   

\section{Non-Debye relaxation models - a brief recollection}\label{sec2}

The non-Debye relaxation processes can be divided into "typical" and ''atypical" phenomena. The words 
"typical" and ''atypical" are referred to Jonscher's universal relaxation law (URL) \cite{AJonscher92} 
which gives the asymptotics of dielectric permittivity $\hat{\varepsilon}(\I\!\omega)$ as the 
function of frequency $\omega$:
   \begin{align}\label{31/03-1}
      \begin{split}
         \hat{\varepsilon}(\I\!\omega) & \propto (\I\!\omega\tau)^{a-1}, \quad \omega\tau \ll 1 \\
         \Delta\hat{\varepsilon}(\I\!\omega) = \varepsilon_{0} - \hat{\varepsilon}(\I\!\omega) & 
				 \propto (\I\!\omega\tau)^{b}, \qquad \omega\tau \gg 1,
      \end{split}
   \end{align}
where the static permittivity $\varepsilon_{0}$ is the limit of $\hat{\varepsilon}(\I\!\omega)$ for 
$\omega\to 0$. The parameters $1-a$ and $b$ belong to the range $(0, 1)$. "Typical" relaxation 
models mean that $b \geq a-1$. Their encompass the HN model and its special cases: the Cole-Cole 
(CC) and the Cole-Davidson (CD) models. An example of "atypical" relaxation is given by the JWS model for which 
$b < a-1$ which also includes the CC model. 

The handful objects investigated in the relaxation physics are spectral functions defined as the ratios of the frequency dependent permittivities $\hat{\phi}(\I\!\omega)=[\hat{\varepsilon}(\I\!\omega) - \varepsilon_{\infty}]/[\varepsilon_{0} -\varepsilon_{\infty}]$. Spectral functions, although being of purely empirical origin, are representable by simple expressions which analytic form comes from fitting the data. If transformed to the time domain  the spectral functions bear the name of the response functions $\phi(\cdot; t)$ introduced either by taking the time derivative of the relaxation function with the sign minus, namely $\phi(\cdot; t) = - \dot{n}(\cdot; t)$ or as the inverse Laplace transform $L^{-1}[\hat{\phi}(\cdot; s),t]$ \cite{footnote3} with $\cdot$ denoting all parameters used to specify the model.  

For the HN relaxation we have
   \begin{multline*}
      \hat{\phi}_{H\!N}(\alpha, \beta; \I\!\omega\tau) = [1 + (\I\!\omega\tau)^{\alpha}]^{-\beta} 
			     \quad \text{and} \\ \phi_{H\!N}(\alpha, \beta; t) = \tau^{-1} (t/\tau)^{\alpha\beta-1} 
					 E_{\alpha, \alpha\beta}^{\,\beta}[-(t/\tau)^{\alpha}]
   \end{multline*}
with $\alpha, \beta\in(0, 1]$ and $E_{\alpha, \nu}^{\mu}(x)$ being the three parameter Mittag-Leffler 
function, see {Appendix B}. For $\beta = 1$ and $\alpha\in(0, 1)$ it reduces to the CC relaxation whereas for 
$\beta\in(0, 1)$ and $\alpha = 1$ it becomes the CD model. If $\alpha = \beta = 1$ then we have the Debye 
relaxation. For the general form of the HN model  the exponents $a$ and $b$ appearing in Jonscher's URL are equal to 
$1 - a = \alpha\beta$ and $b = \alpha$ \cite{RGarrappa16}. 

In the case of JWS model we have
   \begin{multline*}
      \hat{\phi}_{J\!W\!S}(\I\!\omega) = 1 - [1 + (\I\!\omega\tau)^{-\alpha}]^{-\beta} 
			     \quad \text{and} \\ \phi_{J\!W\!S}(t) = \delta(t) - t^{-1} 
					 E_{\alpha, 0}^{\,\beta}[-(t/\tau)^{\alpha}],
   \end{multline*}
where $\delta(t)$ stands for the Dirac $\delta$-function, $\alpha\in(0, 1]$, and $\alpha\beta \leq 1$. 
The exponents in Eq. \eqref{31/03-1} are equal $1 - a = \alpha$ and $b = \alpha\beta$ 
\cite{RGarrappa16}. The JWS model becomes the CC relaxation for $\alpha\in(0, 1)$ and 
$\beta = 1$ and the Debye relaxation for $\alpha = 1$ and $\beta = 1$ analogously as it takes place in the HN model. 

The relaxation function $n(t)$ is connected to the spectral function $\hat{\phi}(\I\!\omega)$ by the the inverse Laplace transform  
   \begin{equation}\label{12/07-1}
      n(t) = L^{-1} [(1-\hat{\phi}(\I\!\omega))/(\I\!\omega); t]\,. 
   \end{equation}
    It implies that the relaxation function  
$n_{HN}(t)$ relevant for the HN model \cite{RGarrappa16, KGorska18} reads
   \begin{equation*}
      n_{H\!N}(t) = 1 - (t/\tau)^{\alpha\beta}\, E_{\alpha, 1 + \alpha\beta}^{\beta}[-(t/\tau)^{\alpha}], 
   \end{equation*}
while for the JWS model it is equal to \cite{RGarrappa16}
   \begin{equation*}
      n_{J\!W\!S}(t) = E_{\alpha, 1}^{\beta}[-(t/\tau)^{\alpha}].
   \end{equation*}
Because of the future use of the CC relaxation function we quote it explicitly
   \begin{equation*}
      n_{CC}(t) = E_{\alpha}[-(t/\tau)^{\alpha}].
   \end{equation*} 
For $\beta = 1$ it comes immediately from $n_{J\!W\!S}(t)$ while for $n_{H\!N}(t)$ emerges after some calculation effort presented in \cite{KGorska18}.

Recall that the transition rate $r(t, \tau) = -\dot{n}(t)/n(t) = 
\phi(t)/n(t)$. The use of asymptotics of $\phi(t)$ and $n(t)$ for $t\to 0$ (given in \cite[Eqs. (3.29), 
(3.30)]{RGarrappa16} as well as in [{\it ibid.}, Eqs. (3.45), (3.46)]) enables one to write 
   \begin{equation}\label{31/05-4}
      r_{H\!N}(t, \tau) \sim \frac{\alpha\beta}{\tau} \frac{(t/\tau)^{\alpha\beta - 1}}
			                       {\Gamma(1 + \alpha\beta) - (t/\tau)^{\alpha\beta}},
   \end{equation}
and
   \begin{equation}\label{31/05-5}
      r_{J\!W\!S}(t, \tau) \sim \frac{\alpha\beta}{\tau} \frac{(t/\tau)^{\alpha-1}}
			                          {\Gamma(1+\alpha) - \beta(t/\tau)^{\alpha}},
   \end{equation}
where $\alpha, \beta \in(0, 1]$. As it has been  mentioned in the Introduction for any $\alpha, \beta \in(0, 1)$ we arrive for $t=0$ at a power-like singularity which vanishes for $\alpha = \beta = 1$, \textit{i.e.}, for the Debye 
relaxation.

\section{{Memory functions} $M(t)$ and $k(t)$}\label{sec3}

{The combination of Eqs. \eqref{15/04-1} and \eqref{12/07-1} taken in the Laplace domain results in 
$\hat{n}(s) = s^{-1} - B\, \hat{M}(s)\, \hat{n}(s)$ or $s\, \hat{n}(s) = 1 - \hat{\phi}(s)$ 
and leads to 
   \[ \hat{M}(s) = B^{-1}\, \hat{\phi}(s)/[s\, \hat{n}(s)], \]
which implies
   \begin{equation} \label{10/07-4}
   		\hat{M}(s) = B^{-1}\,\{ [\hat{\phi}(s)]^{-1} - 1\}^{-1}.
   \end{equation}
The use of Eqs. \eqref{10/07-4} and \eqref{30/05-2} enables us to write
   \[\hat{k}(s) = B\, \hat{n}(s)/\hat{\phi}(s),\] 
and next 
   \begin{equation} \label{10/07-7}
      \hat{k}(s) = B\, s^{-1}\, \{[\hat{\phi}(s)]^{-1} - 1\}. 
   \end{equation}   
The memory functions, as seen from the fact that the Laplace transform $\hat{r}(s, \tau)= L[\phi(t)/n(t); s]$ differs from $\hat{M}(s)$ and $\hat{k}(s)$, are not related in any straightforward way to the transition rate $r(t, \tau)$. More informative is the connection between the {memory functions} $\hat{M}(s)$, $\hat{k}(s)$ and the spectral function 
$\hat{\phi}(s)$. Namely, for large $|\hat{\phi}(s)|^{-1}$ (\textit{i.e.}, small $|\hat{\phi}(s)|$) we can write 
   \begin{equation}\label{15/07-10}
      \hat{M}(s) \propto  \hat{\phi}(s)\,\,\, \text{and}\,\,\, 
			\hat{k}(s) \propto [s \hat{\phi}(s)]^{-1}. 
   \end{equation}
which implies that $\hat{M}(s)$ has the same asymptotics as $\hat{\phi}(s)$,  
whereas $\hat{k}(s)$ behaves as $[s \hat{\phi}(s)]^{-1}$. Thus the asymptotics of 
$\hat{M}(s)$ and $\hat{k}(s)$ may be guessed from the asymptotics of the spectral function 
$\hat{\phi}(s)$ at $\tau |s| \ll 1$ known from the Jonscher URL. 

In the next two subsections we shall focus our interest on finding the {memory functions} $M(t)$ and $k(t)$ for the HN and JWS models.

\subsection*{{Memory functions} for the HN model}

From \eqref{10/07-4} we get the Laplace transformed memory {function} $\hat{M}_{H\!N}(\alpha, 
\beta; s)$ for the HN relaxation 
   \begin{equation*}
      \hat{M}_{H\!N}(\alpha, \beta; s) = B^{-1}\{[1 + (\tau s)^{\alpha}]^{\beta} - 1\}^{-1}. 
   \end{equation*}
Its Laplace inversion is known \cite{AAKhamzin14, CFAERosa15}
   \begin{equation}\label{1/06-2}
      M_{H\!N}(\alpha, \beta; t) = (B t)^{-1}\sum_{r \geq 1} \Big(\frac{t}{\tau}\Big)^{\alpha\beta r} 
			                             E_{\alpha, \alpha\beta r}^{\beta r}[-(t/\tau)^{\alpha}].
   \end{equation} 
Using Eq. \eqref{1/06-2} and expressing $E_{\alpha, \alpha\beta r}^{\beta r}[-(t/\tau)^{\alpha}]$ by the formula Eq. 
\eqref{25/08-2a} we get the integral representation
   \[ M_{H\!N}(\alpha, \beta; t) = B^{-1} \!\!\! \int_{0}^{\infty}\!\!\E^{-u \tau^{-\alpha}}\!\!
			          E_{\beta, 0}\Big[ \Big( \frac{u}{\tau^\alpha}\Big)^\beta\Big] 
								g_{\alpha}(u, t)\, \frac{\D u}u, \]
where the shorthand 
   \[ g_{\alpha}(u, t) = u^{-1/\alpha} g_{\alpha}(t u^{-1/\alpha})\]
has been used to denote the one-sided L\'{e}vy stable distribution 
$g_{\alpha}(x)$ with $\alpha\in(0, 1)$ and $x > 0$ \cite{KAPenson10, HPollard46} (this class of distributions appears also in the KWW relaxation \cite{HPollard48, KGorska12}). The special cases of $M_{H\!N}(\alpha, \beta; t)$ for $\alpha = 1$ and/or $\beta = 1$, \textit{i.e.}, 
the D, CD and CC relaxations, respectively, can be derived either from the series \eqref{1/06-2} or from 
the just given integral form of $M_{H\!N}(\alpha, \beta; t)$. Because the CC relaxation will be used in the sequel we signify the memory {function} $M_{CC}(\alpha; t)$, which coincides with $M_{H\!N}(\alpha, 1; t)$ 
   \begin{equation}\label{15/07-11}
      M_{CC}(\alpha; t) = B^{-1} \frac{\tau^{-\alpha} t^{\alpha-1}}{\Gamma(\alpha)}, \quad \alpha\in(0, 1).
   \end{equation}
The memory {function} $k_{H\!N}(\alpha, \beta; t)$ coupled to $M_{H\!N}(\alpha, \beta; t)$  is 
obtained in the Laplace space from Eq. \eqref{10/07-7} with the spectral function $\hat{\phi}_{H\!N}(\alpha, \beta; s)$ being used. 
That gives
   \begin{equation}\label{3/06-2}
      \hat{k}_{H\!N}(\alpha, \beta; s) = B\,s^{-1}\, [\tau^{\alpha\beta}\,(\tau^{-\alpha} 
			                                 + s^{\alpha})^{\beta} - 1]
   \end{equation}
and in the time domain becomes
   \begin{equation}\label{4/06-1}
      k_{H\!N}(\alpha, \beta; t) = B\, \big\{(\tau/t)^{\alpha\beta} E_{\alpha, 1 
			                           - \alpha\beta}^{-\beta}[-(t/\tau)^{\alpha}] - 1\big\}.
   \end{equation}
To earn the CC relaxation we take $k_{H\!N}(\alpha, 1; t)$. In this case we apply 
the definition of the Mittag-Leffler {polynomials -- displayed in Eq. \eqref{25/08-1a} 
-- according to which 
$E_{\alpha, 1-\alpha}^{-1}(x) = [\Gamma(1-\alpha)]^{-1} - x$}. That allows one to express 
$k_{H\!N}(\alpha, 1; t) \equiv k_{CC}(\alpha; t)$ as
   \begin{equation}\label{8/06-5}
      k_{CC}(\alpha; t) = B\, \frac{(\tau/t)^{\alpha}}{\Gamma(1-\alpha)}, \quad \alpha\in(0, 1).
   \end{equation}
Note that in the Laplace domain {memory functions} $M_{CC}(\alpha;s)$ and $k_{CC}(\alpha;s)$,  $\alpha\in(0, 1)$, take particularly simple forms 
   \begin{equation}\label{20_11_1}
      M_{CC}(\alpha;s)=B^{-1}(\tau s)^{\alpha}; \quad k_{CC}(\alpha; s) = B\tau^{-\alpha}s^{-(1+\alpha)}.
   \end{equation}
\subsection*{{Memory functions} for the JWS model}

For the JWS model we get 
   \[ \hat{M}_{J\!W\!S}(\alpha, \beta; s) = B^{-1}\,[s^{-\alpha\beta} (\tau^{-\alpha} 
	                                        + s^{\alpha})^{\beta} - 1],  \]
which for $\beta = 1$ is tantamount to $\hat{M}_{CC}(\alpha; s)$ whose Laplace transform inversion 
gives $k_{CC}(\alpha; t)$. The inverse Laplace transform of $\hat{M}_{J\!W\!S}(\alpha, \beta; s)$ 
allows one to determine the memory {function} $M_{J\!W\!S}(\alpha, \beta; t)$ in an unique way. It is
   \[ M_{J\!W\!S}(\alpha, \beta; t) = B^{-1}\, \{t^{-1} E_{\alpha, 0}^{-\beta}[-(t/\tau)^{\alpha}] 
			                              - \delta(t)\}. \]
The memory {function} $\hat{k}_{J\!W\!S}(\alpha, \beta; s)$ received from Eq. \eqref{10/07-7} gives
   \begin{equation}\label{16/07-1}
      \hat{k}_{J\!W\!S}(\alpha, \beta; s) = B\, \{s [1 + (\tau s)^{-\alpha}]^{\beta} - s\}^{-1},
   \end{equation}
which for $\beta = 1$ implies $\hat{k}_{CC}(\alpha; s) = \hat{M}_{CC}(\alpha; s)$. To compute 
$k_{J\!W\!S}(\alpha, \beta; t)$ we represent $\hat{k}_{J\!W\!S}(\alpha, \beta; s)$ as $(B x/s)/(1-x)$ where $x = [1 + (s \tau)^{-\alpha}]^{-\beta}$. Assuming that $|x| < 1$ we can express $\hat{k}_{J\!W\!S}(\alpha, \beta; s)$ as
 \begin{equation*}
	    \hat{k}_{J\!W\!S}(\alpha, \beta; s) = B\, \sum_{r\geq 0} \frac{s^{- 1}}
			        {[1 + (s\tau)^{-\alpha}]^{\beta (r + 1)}}
   \end{equation*}
whose Laplace transform reads
\begin{equation*}
k_{J\!W\!S}(\alpha, \beta; t) = B \sum_{r \geq 0} E_{\alpha, 1}^{\beta (r + 1)}[-(t/\tau)^{\alpha}].   
\end{equation*}   
Applying the integral form of $E_{\alpha, 1}^{\beta r}[-(t/\tau)^{\alpha}]$ given by Eq. \eqref{10/11-2a} 
and interchange the order of {integration and summation}, we represent 
$k_{J\!W\!S}(\alpha, \beta; t)$ as follows
   \begin{multline}\label{16/07-12}
      k_{J\!W\!S}(\alpha, \beta; t) = B\, \int_{0}^{\infty} \E^{-\tau^{-\alpha} u} u^{\beta-1} \\ \times L^{-1}[E_{\beta, \beta}(u^{\beta}s^{\alpha\beta}) s^{\alpha\beta-1} 
						 \E^{-u s^{\alpha}}; t] \D u.
   \end{multline}
By virtue of \cite[Eq. (14)]{KGorska20} we write $E_{\beta, \beta}(u^{\beta}s^{\alpha\beta}) = - 
u^{-\beta}s^{-\alpha\beta}E_{-\beta, 0}(u^{-\beta}s^{-\alpha\beta})$, so Eq. \eqref{16/07-12} can be expressed as 
\begin{multline*}
k_{J\!W\!S}(\alpha, \beta; t) = - B\, \int_{0}^{\infty} \E^{-\tau^{-\alpha} u} u^{-1} \\ \times L^{-1}[E_{-\beta, 0}(u^{-\beta}s^{-\alpha\beta}) s^{-1} \E^{-u s^{\alpha}}; t] \D u.
\end{multline*}
Now, using the series form of the two parameter Mittag-Leffler function and once again the integral Eq. \eqref{10/11-2a} we get
\begin{equation}\label{10/11-4}
k_{J\!W\!S}(\alpha, \beta; t) = - B\,\sum_{r\geq 0} E_{\alpha, 1}^{-\beta r}[-(t/\tau)^{\alpha}].
\end{equation}
In the Laplace domain it reads
\begin{equation}\label{10/11-5}
\hat{k}_{J\!W\!S}(\alpha, \beta; s) = - \frac{B}{s}\,\sum_{r\geq 0} \frac{1}{[1 + (s \tau)^{-\alpha}]^{-\beta r}}.
\end{equation}
which can be calculated for $|[1 + (s \tau)^{-\alpha}]^{\beta}| < 1$.  So we obtain Eq. \eqref{16/07-1}. 
In this way we apply the procedure presented in \cite{KGorska19, KGorska20}. Accordingly, the right hand 
side expression in Eq. \eqref{16/07-1} can be expressed in the series form in two different ways which 
gives the solution under assumption $|[1 + (\tau s)^{-\alpha}]^{\beta}| > 1$ or 
$|[1 + (\tau s)^{-\alpha}]^{\beta}| < 1$. Above, we have shown that in both these regions the solutions 
are equivalent. The CC relaxation model is also obtained from Eq. \eqref{16/07-12} by the 
{appropriate} setting $E_{1, 0}(u s^{\alpha}) = u s^{\alpha} \exp(u s^{\alpha})$. 
That enables one to write Eq. \eqref{16/07-12} in the form: 
   \[ k_{J\!W\!S}(\alpha, 1; t) = B \int_{0}^{\infty}\E^{-\tau^{-\alpha} u} u^{-1} 
                                  L^{-1}[u s^{\alpha-1}; t] \D u, \]
which boils down to Eq. \eqref{8/06-5}.

\section{The time evolution equations}\label{sec4}
\subsection*{Two types of the time smearing}

To show that two types of the time smearing applied to Eq. \eqref{28/05-1} lead, under the 
condition \eqref{22/07-1}, to the same result we begin with taking the time derivative of Eq. \eqref{15/04-1}, {\it viz.}
   \begin{align*}
         \dot{n}(t) & = -B \frac{\D}{\D t} \int_{0}^{t} M(t - \xi) n(\xi) \D\xi \\
                    & = -B \frac{\D}{\D t} \int_{0}^{t} M(\eta) n(t-\eta) \D\eta .
   \end{align*}
In turn, thanks to the standard procedure involving Leibniz rule, replacement of the derivative 
$\frac{\D}{\D t}\rightarrow -\frac{\D}{\D\eta}$ and integration by parts, we transform the latter into
   \begin{align}\label{21/05-4}
      - \dot{n}(t) &= B M(t) + B \int_{0}^{t} M(\eta) \,\dot{n}(t - \eta) \D\eta \notag \\
                   &= B M(t) + B \int_{0}^{t} M(t-\xi) \,\dot{n}(\xi)  \D\xi\,,
   \end{align}
which is in fact the integral equation for the response function $\phi(t)$.  
Multiplying Eq. \eqref{21/05-4} by $k(T-t)$ and integrating it with respect to $t$ on $[0,T]$ we 
arrive at the equation which right-hand side involves the time smeared $\dot{n}(t)$. Consequently,
   \begin{multline}\label{21/05-5}
         \int_{0}^{T} k(T - t) \dot{n}(t) \D t = -B \int_{0}^{T}\!\!k(T-t) M(t) \D t  \\ - B \int_{0}^{T} \Bigg[\int_{0}^{t} k(T-t) M(t-\xi) \dot{n}(\xi) \D\xi\Bigg] \D t.
   \end{multline}
{The integration domain of the double integral is of right triangle form so, 
setting $T-t = u$ we can rewrite the second line of Eq. \eqref{21/05-5} into}
   \begin{multline*}
      - B\int_{0}^{T} \Bigg[\int_{0}^{t} k(T-t) M(t-\xi) \dot{n}(\xi) \D\xi\Bigg] \D t \\ =
      - B\int_{0}^{T} \Bigg[\int_{0}^{T-\xi} k(u) M(T-\xi - u) \D u\Bigg] \dot{n}(\xi) \D\xi. 
   \end{multline*}
If the memory kernels $M(t)$ and $k(t)$ satisfy Eq. \eqref{22/07-1} then the RHS above equals $-Bn(t) + B$ and we get Eq. \eqref{29/05-3}. 
Thus, we can claim that the condition \eqref{30/05-2} makes Eq. \eqref{29/05-3} equivalent to 
Eq. \eqref{15/04-1}. We can also conclude that the smearing of $r(t, \tau) n(t)$ can be exchanged 
by the smearing of the first time derivative $\dot n(t)$ as it has been shown in \cite{KGorska20}.

\subsection*{The Cole-Cole model: Riemann-Liouville integrals and Caputo derivatives}

The simplest example which illustrates the meaning and practical usefulness of the relation between Eqs. \eqref{15/04-1} and \eqref{29/05-3} is given by the CC model.  
Eq. \eqref{15/04-1} with $M_{CC}(\alpha; t)$ of Eq.\eqref{15/07-11} reads
   \begin{equation}\label{19/08-1}
      n_{CC}(t) = 1 - \tau^{-1} (I^{\alpha}\, n_{CC})(t),
   \end{equation}
where 
   \begin{equation*}
      (I^{\alpha} f)(t) = \frac1{\Gamma(\alpha)} \int_0^t f(\xi) (t - \xi)^{\alpha - 1} \D\xi, 
			                    \quad \alpha > 0
   \end{equation*}
is the Riemann--Liouville fractional integral equivalently defined as the negative fractional derivative in 
the Riemann--Liouville sense, namely $I^\alpha_0 = D^{-\alpha}_0$. Taking the Riemann-Liouville fractional derivative 
$(D^{\alpha} f)(x) = (\frac{\D}{\D x} I^{1-\alpha} f)(x)$ of both sides 
of Eq. \eqref{19/08-1} we get
   \begin{equation}\label{19/08-3}
      (D^{\alpha}\, n_{CC})(t) = D^{\alpha} 1 - \tau^{-\alpha} (D^{\alpha} I^{\alpha}\, n_{CC})(t), 
   \end{equation}
which is nothing else than Eqs. (2.141) and (2.106) of \cite{IPodlubny99} for $\alpha\in(0, 1)$. Standard relation 
$D^{\alpha} C = C\, x^{-\alpha}/\Gamma(1-\alpha)$ valid for any constant $C$ and the semigroup property$(D^{\alpha} D^{-\alpha} f)(x) 
= f(x)$ allow one to rewrite Eq. \eqref{19/08-3} as 
   \begin{equation}\label{20/08-1}
      (D^{\alpha}\, n_{CC})(t) - \frac{t^{-\alpha}}{\Gamma(1-\alpha)} = - \tau^{-\alpha} n_{CC} (t).
   \end{equation}
Two terms in the left hand side of Eq. \eqref{20/08-1}, with $\alpha\in(0, 1)$,  reduce to
the fractional derivative in the Caputo sense $({^{C\!}D^{\alpha}}\, n_{CC})(t)$ which 
coincides with $ (I^{1-\alpha} \frac{\D}{\D t}\, n_{CC})(t)$, also with $\alpha\in(0, 1)$. Hence, the integral equation 
\eqref{19/08-1} becomes the integro--differential equation 
   \begin{equation}\label{20/08-2}
      {^{C\!}D^{\alpha}} n_{CC}(t) = - \tau^{-\alpha} n_{CC} (t),
   \end{equation} 
which we recognize as Eq. \eqref{29/05-3} with the memory {function} $k_{CC}(\alpha; t)$ given by Eq. \eqref{8/06-5}. 
Thus our approach reproduces classical forms  of the evolution equations governing the CC model: Eq. \eqref{19/08-1} is equivalent to \cite[Eq. (3.7.43)]{RGorenflo14}, Eq. \eqref{20/08-1} 
is analogous to \cite[Eq. (3.7.49)]{RGorenflo14}, and Eq. \eqref{20/08-2} is the same as 
\cite[Eq. (3.10)]{RGarrappa16}.

\subsection*{The Havriliak-Negami and Jurlewicz-Weron-Stanislavsky models: more complicated pseudo-operators enter the game}

As remarked in the first subsection of Sec. \ref{sec4} it is natural to treat Eqs. \eqref{15/04-1} and \eqref{29/05-3} as equivalent. Thus, to get information on the process, it is enough to find and solve one of these equations. Knowing the {memory functions} $k_{H\!N}(\alpha, \beta; t)$ and 
$M_{J\!W\!S}(\alpha, \beta; t)$ we can get relevant forms of Eqs. \eqref{15/04-1} and \eqref{29/05-3} without tedious 
calculations involving operator and power series methods {\cite{RGarrappa16, RNigmatullin, AStanislavsky16}}. If we focus our interest on the HN relaxation then we take Eq. 
\eqref{29/05-3} with memory {function} $k_{H\!N}(\alpha, \beta; t)$. It immediately yields the equation proposed in  
\cite[Eq. (3.40)]{RGarrappa16}:
   \begin{equation}\label{24/08-2}
      {^{\rm{C}}(D^{\alpha} + \tau^{-\alpha})^{\beta}}\, n(t) = - \tau^{-\alpha\beta},
   \end{equation}
where the pseudo--operator on the left hand side belongs to the class of Prabhakar--like integral operators 
and for the considered case it is defined as 
\cite[Eq. (B.23)]{RGarrappa16}  
   \begin{multline}\label{24/08-3}
      {^{\rm{C}}(D^{\alpha} + \tau^{-\alpha})^{\beta}} n(t) \\
			   = \int_{0}^{t} (t - \xi)^{-\alpha\beta} E_{\alpha, 1 - \alpha\beta}^{-\beta}
				   \Big[-\dfrac{(t - \xi)^{\alpha}}{\tau^{\alpha}}\Big]\, \dot{n}(\xi) \D\xi.
   \end{multline}
Note that we are keeping the notation of \cite[Appendix B]{RGarrappa16} and it has to be pointed out 
here that the pseudo--operator \eqref{24/08-3} must be distinguished from the pseudo--operator  
$({^{C\!}D}^{\alpha} + \tau^{-\alpha})^{\beta}$ involving the Caputo derivative (also considered in 
the paper \cite{RGarrappa16}). {This is evidently seen from fact that using Eq. \eqref{25/08-1a} it can 
be concluded} ${^{\rm{C\!}}(D^{\alpha} + \tau^{-\alpha})} \,n(t) = ( {^{C\!}D}^{\alpha} 
+ \tau^{-\alpha})n(t) - \tau^{-\alpha}$.

Looking for the equation governing the JWS model we take Eq. \eqref{15/04-1} with the {memory kernel} 
$M_{J\!W\!S}(\alpha, \beta; t)$. That gives
   \begin{equation}\label{24/08-5}
      \int_0^t (t-\xi)^{-1} E_{\alpha, 0}^{-\beta}[-\tau^{-\alpha}(t-\xi)^{\alpha}]\, n(\xi) \D\xi = 1,
\end{equation}
where we have used the sifting (or sampling) property of the Dirac 
$\delta$--function, that is $\int_0^t \delta(t - \xi) n(\xi) \D\xi = n(t)$. Thus}, the integrand in the 
formula Eq. \eqref{24/08-5} can be rewritten due {to} the substitution 
   \[ t^{-1} E_{\alpha, 0}^{-\beta}(-\tau^{-\alpha} t^{\alpha}) = D^{1-\alpha\beta}[t^{-\alpha\beta} 
	           E_{\alpha, 1 - \alpha\beta}^{-\beta}(-\tau^{-\alpha}t^{\alpha})]. \]
quoted in  Eq. \eqref{24/08-1a}. Thereafter, we take the Riemann-Liouville fractional derivative 
$D^{\alpha\beta}$ of both sides of the resulting equation and make use of the semigroup 
property $(D^{\mu}D^{\nu} f)(x) = (D^{\mu+\nu} f)(x)$ \cite[Eq. (2.127)]{IPodlubny99} taken for $\mu=\alpha\beta$ and $\nu=1-\alpha\beta$. Thus, we obtain
\begin{equation}\label{21/03-1}
 \int_0^t \frac{\D}{\D t} \{(t-\xi)^{-\alpha\beta} E_{\alpha, 1 - \alpha\beta}^{-\beta}
	             [-\tau^{-\alpha}(t-\xi)^{\alpha}]\} n(\xi)\D\xi = D^{\alpha\beta} 1.
\end{equation}	             
Interchange (in a legitimate way) of the order of integration and derivation enables us to represent the left-hand side of the equation above as the action of the pseudo--operator $(D^{\alpha} + \tau^{-\alpha})^{\beta}$, {\it viz.}
   \begin{multline*}
       \frac{\D}{\D t}\int_{0}^{t} (t-\xi)^{-\alpha\beta} 
			     E_{\alpha, 1 - \alpha\beta}^{-\beta}\Big[-\frac{(t-\xi)^{\alpha}}{\tau^{\alpha}}\Big]\, 
									n(\xi)\, \D\xi \\ = (D^{\alpha} + \tau^{-\alpha})^{\beta}n(t).
   \end{multline*}
{Using this Eq. \eqref{21/03-1} takes the form of kinetic equation
 \begin{equation}\label{21/03-2}
 (D^{\alpha} + \tau^{-\alpha})^{\beta} n(t) = \frac{t^{-\alpha\beta}}{\Gamma(1-\alpha\beta)},
 \end{equation}
which should be completed with a suitable initial condition. Eq. \eqref{21/03-2}, in the context of the JWS model, first appeared in \cite[Eq. (4.3)]{AStanislavsky16} being justified by under- and overshooting subordination technique applied for the anomalous diffusion \cite{AStanislavsky10}. Eq. \eqref{21/03-2} was also obtained using the integral transform methods in \cite{RGarrappa16}. } 

\section{Towards multichannel description of relaxation phenomena}\label{sec5}

Hitherto, we have considered models describing the relaxation phenomena assumed to be underpinned by a single stochastic process $U$ characterized by the Laplace exponent $\Psi(s)$. The complete Bernstein character of $\hat{\Psi}(s)$ allows to introduce a pair of coupled {memory functions} related to this process, namely $M(t)$ and $k(t)$, whose Laplace transforms are given as $\hat{M}(s) = [\hat{\Psi}(s)]^{-1}$ and $\hat{k}(s) = \hat{\Psi}(s)/s$ and consequently satisfy a constraint $\hat{M}(s)\hat{k}(s) = s^{-1}$. The latter leads to the conclusion that integral or integro-differential equations Eqs. \eqref{15/04-1} and \eqref{29/05-3} which even if formulated in a two-fold way emerge from the same stochastic process $U$ and describe the same physical reality. 

In what follows we are going to pay attention on an extension of the presented approach which sheds light on  relaxation processes assumed to be described by effective spectral functions composed of more elementary ones, say $\hat{\phi}_j(s), \, j=1,2$ being e.g. any of the standard non--Debye patterns. Each such component, if treated separately, leads to the completely monotone memory {function} and is related to its own Laplace exponent $\hat{\Psi}_j(s)$. 

In the framework of the linear response approach we may consider a simple model for which the effective spectral function is represented {in the form of a linear combination $\hat{\phi}(s) = 
C_1\hat{\phi}_{1}(s) + C_2 \hat{\phi}_{2}(s)$, with positive coefficients $C_1$ and $C_2$} next put equal 
to $1$ to simplify the exposition. Obviously $\hat{\phi}(s)$ is CMF and may be written as 
\[ \hat{\phi}(s) = [1+\Psi_{1}(s)]^{-1} + [1+\Psi_{2}(s)]^{-1}\,.\]
The following question immediately arises - is it possible  to relate $\hat{\phi}(s)$ with a single 
process arisen as a cumulative effect of {processes $U_1$ and $U_2$, say?} To answer such a question we should determine the 
new Laplace exponent $\hat{\Psi}(s)$ which would satisfy $\hat{\phi}(s) = [1+\hat{\Psi}(s)]^{-1}$ and 
be a CBF. We have
\begin{align}\label{18_11_2}
&\hat{\phi}(s) = \hat{\phi}_{1}(s) + \hat{\phi}_{2}(s) 
               = \dfrac{1+\hat{\Psi}_{1}(s)+1+\hat{\Psi}_{2}(s)}
							   {[1+\hat{\Psi}_{1}(s)][1+\hat{\Psi}_{2}(s)]}\notag \\
              &\quad = \dfrac{1+\displaystyle{\frac1{1+\hat{\Psi}_{1}(s)+\hat{\Psi}_{2}(s)}}}
							   {1+\displaystyle{\frac{{\hat\Psi}_{1}(s)\hat{\Psi}_{2}(s)}{1+\hat{\Psi}_{1}(s)
							 + \hat{\Psi}_{2}(s)}}} = \displaystyle{\frac{1}{1+\hat{\Psi}(s)}}. 
\end{align}
From that we conclude $\hat{\Psi}(s) = [1 - \hat{\phi}(s)]/\{1 - [1 - \hat{\phi}(s)]\}$, which after 
substituting Eq. \eqref{18_11_2} {\it mutatis mutandis} takes the series expansion form
   \begin{equation} \label{2021_02_02} 
	    \hat{\Psi}(s) = \sum_{r\geq 0} \left(1- \displaystyle{\frac{1+\displaystyle{\frac1{1+\hat\Psi_1(s) 
	                  + \hat\Psi_2(s)}}}{1+\displaystyle{\frac{{\hat\Psi}_1(s)\hat{\Psi}_2(s)}
										  {1+\hat{\Psi}_1(s)+\hat{\Psi}_2(s)}}}}\right)^{r+1}. 
	 \end{equation}
{We point out that this series expansion is well--defined, being for $x = \hat \Phi_1(s), 
y = \hat \Phi_2(s)$ the estimate 
   \[ \Big| \dfrac{x y -1}{(1+x)(1+y)}\Big|<1, \qquad x, y>0  \]
valid. This implies the geometric series expansion \eqref{2021_02_02}. On the other hand} 
$\hat{\Psi}(s)$ is CBF when it is the series of CBFs and/or the first term in the 
series is completely monotone. For any CBF $\hat{\Psi}_1(s)$ and $\hat{\Psi}_2(s)$ the numerator of 
this expression is CMF as a reciprocal of CBF. 

To guarantee complete monotonicity 
of the whole expression it will be sufficient to show that its denominator is a CBF which comes from 
the property that any fraction CMF/CBF is CBF. Obviously $1+\hat{\Psi}_1(s)+\hat{\Psi}_2(s)$ is 
the complete Bernstein function which helps us to analyse analogous property of $\hat{\Psi}_1(s)
\hat{\Psi}_2(s)/[1+\hat{\Psi}_1(s)+\hat{\Psi}_2(s)]$. Using \cite[Proposition 7.1] {RSchilling10} we 
learn that the sufficient condition for the latter to be CBF is $\hat{\Psi}_1(s)\hat{\Psi}_2(s)=s$. 
Recalling the analytic forms of $\hat{\Psi}$'s relevant for the standard non-Debye relaxation patterns 
   \begin{center}
      \begin{tabular}{|c|c|c|} \hline
	       & $\hat{M}(s)$	& $\hat{\Psi}(s)$\\ \hline
       CC& $B^{-1}\{(\tau s)^{\alpha}]\}^{-1}$  & $B (\tau s)^{\alpha}$ \\ \hline
       HN& $B^{-1}\{[1 + (\tau s)^{\alpha}]^{\beta} - 1\}^{-1}$  
			          & $B s\{[1 + (\tau s)^{\alpha}]^{\beta} - 1$\} \\ \hline
      JWS& $ B^{-1}[1 + (\tau s)^{-\alpha}]^{\beta} - 1$ 
			          & $B s\{[1 + (\tau s)^{-\alpha}]^{\beta} - 1\}$\\ \hline
      \end{tabular}
   \end{center} 
we see that this condition may be satisfied only for $\hat{\Psi}_1(s)=s^{\alpha}$, $\hat{\Psi}_1(s)
=s^{\beta}$, which fulfill the additional constraint $\alpha + \beta =1$. Therefore, we conclude that our simple ``two channel'' model, if based on the sum of Cole-Cole  patterns, share properties of standard ``one channel'' models and may be proposed as a prospective challenger for description phenomena which exhibit more than one characteristic time or go beyond the Jonscher URL. This confirms theoretical results of \cite{AStanislavsky20} and also justifies approach proposed in \cite{Konieczny} to interpret experimental data obtained in studies of magnetic relaxation and magneto-calorimetric effect.

\section{Conclusions}\label{sec6} 

We have considered the HN and JWS relaxation patterns starting from the basic assumption that relaxation models are governed by integral or integro-differential equations with kernels which mimic memory effects and in such a way influence the time behaviour of relaxing systems. Equations been studied are of the Volterra type and involve the time non-localities, expressed either through the integral operator with the time smeared integral kernel denoted by $M$ or contain the generalized (fractional) differential operator involving the time non-local kernel $k$. If the kernels $M$ and $k$ satisfy the Sonine condition then both equations yield equivalent results for the relaxation function $n(t)$ has been looked for. 

Simultaneously,  such derived equations stem from the analysis of the stochastic processes which govern the relaxation and mathematically are described by infinite divisible distributions. It implies that the {memory functions} $M$ and $k$ responsible for behavior of the physical system are uniquely connected with the Laplace exponent which is mathematical characterization of the underlying process. Our approach merges dynamical and statistical/stochastic aspects of the relaxation phenomena through the positive-valued functions which properties guarantee that both physical as well as mathematical requirements of the theory are satisfied.   

The simplest illustration relevant for our construction are equations relevant for the Cole-Cole relaxation pattern. They contain either the Riemann--Liouville fractional integral \eqref{19/08-1} or the fractional derivatives in the Caputo sense \eqref{20/08-2} and are solvable by the Mittag-Leffler function. Physically the first type of these equations describes the smearing of $r_{CC}(t, \tau) n_{CC}(t)$ and another one characterizes the smearing of $\dot{n}_{CC}(t)$. 

Analogical, however much complicated, situation we meet for the HN and JWS models. Nevetheless pseudo--operators traditionally used in their description may be treated as connected with the same types of the time-smearing: the pseudo--operator ${^{\rm C}(D^{\alpha} + \tau^{-\alpha})^{\beta}}n(t)$ reflects the time-smearing $\dot{n}(t)$ whereas $(D^{\alpha} + \tau^{-\alpha})^{\beta} n(t)$ is related 
to the time-smeared product of $n(t)$ and the transition rate $r(t, \tau)$. We are convinced that our approach provides us with mathematically well-defined integral or integro-differential counterparts of pseudo--differential operators and will be helpful to clarify this duality. 

Resumming the paper  - the presented approach enables us to explain, in the framework of one scheme, the origin of the time evolution equations used to describe the relaxation phenomena. This flows out from  two sources:  first, the relaxation phenomena are rooted in stochastic processes which general (mathematical) properties govern their physical properties and second, evolution equations describing dynamics of the relaxation phenomena involve memory effects encoded by functions directly related to objects which characterize underlying stochastic processes. Moreover, our scheme opens possibilities to develop and verify methods leading to the consistent description of multichannel processes.

\section*{Acknowledgments}

{The authors express the gratitude to the anonymous referee for turning their attention to Ref. \cite{AStanislavsky16} which Sec. 4 clarifies the role of pseudo-operators appearing in description of JWS and HN models.}

K.G. and A.H. were supported by the Polish National Center for Science (NCN) research grant OPUS12 
no. UMO-2016/23/B/ST3/01714. The research of T.K.P. has been supported in part by the University 
of Rijeka, Croatia, under the project {\tt uniri-pr-prirod-19-16}.  

\appendix

\section{Classes of positive functions: completely monotone, Bernstein and completely Bernstein}\label{appA}

The completely monotone functions (CMFs) $F(x)$ are non-negative functions on $\mathbb R_>$ whose 
all derivatives exist and  alternate, \textit{i.e.}, 
   \[ (-1)^{n} F^{(n)}(x) \geq 0, \quad n = 0, 1, \ldots, \]
where $F^{(n)}(x) = \D^{\,n} F(x)/ \D x^{n}$. According to the Bernstein theorem \cite{RSchilling10}: 
$x\in [0,\infty)\rightarrow F(x) \in \textrm{CMF}$ {\bf iff}  
   \begin{equation}\label{17/06-1}
      F(x) = \int_{0}^{\infty} \exp(-x t) g(t)\!\D t 
   \end{equation} 
where $g(t)\ge 0$ for $t\in [0,\infty)$. Eq. \eqref{17/06-1} uniquely connects the CMF with the 
non-negative integrable function. It should be also remembered that the product and convex sum of two CMFs 
is a CMF.

The Bernstein functions (BF) are non-negative functions on $\mathbb R_> $ having CMF derivative 
\cite{RSchilling10}: $h(s) > 0$ is a BF if
   \[ (-1)^{n-1} h^{(n)}(s) \geq 0, \quad n = 1, 2, \ldots. \]
Moreover, $h(s)$ is a BF iff it admits the representation
   \begin{equation*}
      h(s) = a + b s + \int_{0}^{\infty} (1-\E^{-s \xi}) \D\mu(\xi),
   \end{equation*}
where $a, b \geq 0$ and $\mu$ (the L\'{e}vy measure) is a positive measure on $(0, \infty)$ satisfying 
$\int_{\mathbb R_>}\xi/(\xi + 1)\D\mu(\xi) < \infty$, see \cite[Theorem 2.6]{Berg} or 
\cite[Theorem 3.2]{RSchilling10}. The following properties of the BF deserve to be mentioned: 

\begin{itemize}
\item[(BF1)] the convex sum of two BFs is another BF; 
\item[(BF2)] the composition of BFs is also a BF, i.e $h_{1}(h_{2}(s))$ is a BF for all $h_{1}$ 
and $h_{2}$ being BFs; 
\item[(BF3)] the composition of a CMF and a BF is another CMF, i.e. $F(h(s))$ is a CMF.
\end{itemize}

The complete Bernstein functions (CBF) \cite{RSchilling10} are defined: $c(s)$ is a CBF, $s > 0$, 
if $c(s)/s$ is the Laplace transform of a CMF restricted to the positive semiaxis or, equivalently, 
the Stieltjes transform of a positive function restricted to this domain - this comes from the fact 
all Stieltjes functions are completely monotone. CBFs is a subclass of BFs so the properties 
(BF1)-(BF3) can be also rewritten for CBFs. Thus, 

\begin{itemize}
\item[(CBF1)] the convex sum of two CBFs is another CBF; 
\item[(CBF2)] the composition of CBFs is a CBF; 
\item[(CBF3)] the composition of a CMF and a CBF is another CMF;
\end{itemize}

are complemented by

\begin{itemize}
\item[(CBF4)] if $c(s)$ is a CBF then $s/c(s)$ is a CBF; \\ [-1.6 \baselineskip]
\item[(CBF5)] if $\alpha, \beta\in(0, 1)$ such that $\alpha+\beta\leq 1$ then $c_{1}^{\alpha} 
\cdot c_{2}^{\beta}$ is a CBF for all $c_{1}$ and $c_{2}$ being CBFs.
\end{itemize}

For our purposes the most important is the property CBF4.

\section{Three parameter Mittag-Leffler function}\label{appB}

The three parameter Mittag-Leffler function is defined through the power series 
   \begin{equation}\label{24/03-a2}
      E_{\alpha, \nu}^{\mu}(x) = \sum_{r\geq 0} \frac{(\mu)_{r} z^{r}}{r! \Gamma(\nu + \alpha r)},
   \end{equation} 
where $\Re(\alpha), \Re(\nu), \Re(\mu) > 0$ and $z \in \mathbb{C}$, $(\mu)_{r}$ denotes 
the Pochhammer symbol (raising factorial) equal to $\Gamma(\mu + r)/\Gamma(\mu) = \mu (\mu + 1)\ldots 
(\mu + r-1); r \in \mathbb N_0$. The Pochhammer symbol for $\mu = 1$ is equal to $r!$ and 
\eqref{24/03-a2} depends on  two parameters $\alpha$ and $\nu$ only. This case is named the two 
parameter Mittag-Leffler (Wiman) function and it is quoted as $E_{\alpha, \nu}(z) = E^1_{\alpha, \nu}(z)$. 
For $\mu = \nu = 1$ Eq. \eqref{24/03-a2} reduces to the one parameter (standard) Mittag-Leffler function 
$E_{\alpha}(z) = E_{\alpha, 1}^1(z)$ which generalizes the exponential function. For 
$\alpha,\beta,\gamma >0$, $\beta - \alpha\gamma>0$ and $z \in \mathbb {R}_{<}$  the three, two and 
one parameter Mittag-Leffler functions are completely monotone. They find numerous applications in 
the relaxation theory coming from the frequently met shape of their Laplace transform  
   \begin{equation}\label{24/03-a1}
      \mathcal{L}[t^{\nu - 1} E_{\alpha, \nu}^\mu(\lambda t^\alpha); s] = 
			         s^{-\nu}(1-\lambda s^{-\alpha})^{-\mu}\,
   \end{equation} 
for $\Re(\nu), \Re(s) > 0$, $|s| > |\lambda|^{1/\Re(\alpha)}$ \cite{TRPrabhakar69}. The fractional 
derivative in Riemann--Liouville sense of $t^{\nu-1} E_{\alpha, \nu}^{\mu}(\lambda t^\alpha)$ 
yields Eq. (5.1.34) of \cite{RGorenflo14} namely
   \begin{equation}\label{24/08-1a}
      \{D^{\beta} [t^{\nu-1}E_{\alpha, \nu}^{\mu}(\lambda t^{\alpha})]\}(x) 
			     = x^{\nu-\beta - 1} E_{\alpha, \nu-\beta}^{\mu}(\lambda x^{\alpha}).
   \end{equation}
The general form of the integral representation of $E_{\alpha, \nu}^{\mu}(-\lambda t^{\alpha})$ is given by \cite[Eq. (15)]{KGorska20}. In the paper we need only two cases, namely
\begin{equation}\label{10/11-2a}
E_{\alpha, 1}^{\gamma}(-\lambda t^{\alpha}) = \frac{1}{\Gamma(\gamma)} \int_{0}^{\infty} \E^{-\lambda u} u^{\gamma - 1} \mathcal{L}^{-1}[s^{\alpha\gamma -1} \E^{-u s^{\alpha}}; t] \D u
\end{equation}
and derived in \cite[Eq. (11)]{KGorska18} or \cite[Eq. (17)]{KGorska20}
   \begin{equation}\label{25/08-2a}
      E_{\alpha, \alpha\mu}^{\mu}(-\lambda t^{\alpha}) = t^{1-\alpha\mu} \int_{0}^{\infty} 
			           \E^{-\lambda u} u^{\mu - 1} g_{\alpha}(u, t),
   \end{equation}
where $g_{\alpha}(u, t) = \mathcal{L}^{-1}[\E^{-u s^{\alpha}}; t]$. Moreover, $g_{\alpha}(u, t) = u^{-1/\alpha} g_{\alpha}(1, t u^{-1/\alpha}) = u^{-1/\alpha} g_{\alpha}(t u^{-1/\alpha})$ and $g_{\alpha}(\sigma)$ with $\sigma > 0$ and $\alpha\in(0, 1)$ is the 
one-sided L\'{e}vy stable distribution \cite{KAPenson10, HPollard46}. From the definition 
\eqref{24/03-a2} we can also derive the Mittag-Leffler polynomials \cite[Eq. (22)]{KGorska20}
   \begin{equation}\label{25/08-1a}
      E^{-n}_{\alpha, \nu}(x) = \sum_{r=0}^{n} \binom{n}{r} \frac{(-x)^{r}}{\Gamma(\nu + \alpha r)}, 
			                          \quad \alpha, \nu > 0,
   \end{equation}
which also satisfied properties \eqref{24/03-a1} and \eqref{24/08-1a}.



\begin{thebibliography}{99}

\bibitem{Berg}
{Ch. Berg, {\em Stieltjes-Pick-Bernstein-Schoenberg and their connection to completely monotonicity}, 
in "Positive Define Functions: From Schoenberg to Space-Time challenges", edited by J. Mateu and 
E. Porcu, (Dep. Math. of Univ. Jaume I, Castellon, Spain, 2008).}

\bibitem{Boettcher} C. J. F. B\"ottcher and P. Bordewijk, \textit{Theory of Electric Polarization}, Elsevier, Amsterdam 1978

\bibitem{Konieczny} M. Fitta, R. Pe{\l}ka, P. Konieczny, M. Ba{\l}anda, {\it Multifunctional molecular 
magnets: magnetocaloric effect in octacyanometallates}, Crystals {\bf 9} (2019) 9.

\bibitem{RGarrappa16} R. Garrappa, F. Mainardi, and G. Maione, {\em Models of dielectric relaxation 
based on completely monotone functions}, Frac. Calc. Appl. Anal. {\bf 19}(5) (2016) 1105--1160; 
corrected version available in {\tt arXiv: 1611.04028}

\bibitem{Gloeckle} W. G. Gl\"ockle and T. F. Nonnenmacher, {\em Fox function representation of 
non-Debye relaxation processes}, J. Stat. Phys. {\bf 71} (1993) 741--757.

\bibitem{RGorenflo14} R. Gorenflo, A. A. Kilbas, F. Mainardi, and S. V. Rogosin, "Mittag-Leffler 
Functions, Related Topics and Applications" (Springer, Berlin, 2014).

\bibitem{KGorska12} K. G\'{o}rska, K. A. Penson, G. Dattoli, and G. H. E. Duchamp, {\em Operator solutions for fractional Fokker--Planck equations}, Phys. Rev. E {\bf 85} (2012) 031138.

\bibitem{KGorska18} K. G\'{o}rska, A. Horzela, {\L}. Bratek, K. A. Penson, and G. Dattoli, 
{\em The Havriliak-Negami relaxation and its relatives: the response, relaxation and probability 
density functions}, J. Phys. A: Math. Theor. {\bf 51} (2018) 135202.

\bibitem{KGorska19} K. G\'{o}rska, A. Horzela, and T. K. Pog\'{a}ny, {\em A note on the article 
"Anomalous relaxation model based on the fractional derivative with a Prabhakar-like kernel" 
[Z. Angew. Math. Phys. (2019) 70: 42]}, Z. Angew. Math. Phys. {\bf 70} (2019) 141.

\bibitem{KGorska20} K. G\'{o}rska and A. Horzela, {\em The Volterra type equations related to the 
non-Debye relaxation}, Commun. Nonlinear Sci. Numer. Simulat. {\bf 85} (2020) 105246.

\bibitem{KGorska20a} K. G\'orska, A. Horzela, E. K. Lenzi, G. Pagnini, and T. Sandev, {\em Generalized Cattaneo (telegrapher's) equation in modeling anomalous diffusion phenomena}, Phys. Rev. E {\bf 102} (2020) 022128.

\bibitem{AHanyga20} A. Hanyga, {\em A comment on a controversial issue: a generalized fractional 
derivative cannot have a regular kernel}, Fract. Calcul. Appl. Anal. {\bf 23} (2020) 211.

\bibitem{RHilfer02a} R. Hilfer, {\em $H$-function representations for stretched exponential relaxation 
and non-Debye susceptibilities in glassy systems}, Phys. Rev. E {\bf 65} (2002) 061510.

\bibitem{AJonscher92} A. K. Jonscher, {\em The universal dielectric response and its physical 
significance}, IEEE Transactions on Electrical Insulation {\bf 27} (1992) 407.

\bibitem{AAKhamzin14} A. A. Khamzin, R. R. Nigmatullin, and I. I. Popov, {\em Justification of 
the empirical laws of the anomalous dielectric relaxation in the framework of the memory function 
formalism}, Fract. Calc. Appl. Anal. {\bf 17} (2014) 247.

\bibitem{ANKochubei11} A. N. Kochubei, {\em General fractional calculus, evolution equations, and 
renewal processes}, Integr. Equations. Oper. Theory {\bf 71} (2011) 583--600.

\bibitem{KremerSchoenhals} F. Kremer and A. Sch\"onhals, \textit{Broadband Dielectric Spectroscopy}, Berlin Heidelberg  Springer Verlag 2003.

\bibitem{Mainardi18} F. Mainardi, {\em A note on the equivalence of fractional equations to 
differential equations with varying coefficients}, Mathematics {\bf 6} (2018), Article ID=8, 5pp.

\bibitem{RNigmatullin} R. R. Nigmatullin and Ya. E. Ryabov, {\em Cole-Davidson dielectric relaxation as a self-similar relaxation process}, Phys. Stat. Sol. {\bf 39} (1997) 101--105.


\bibitem{KAPenson10} K. A. Penson and K. G\'{o}rska, {\em Exact and explicit probability densities for 
one-sided L\'{e}vy stable distributions}, Phys. Rev. Lett. {\bf 105} (2010) 210604.

\bibitem{HPollard46} H. Pollard, {\em The representation of $\E^{-x^{\alpha}}$ as a Laplace 
integral}, Bull. Amer. Math. Soc. {\bf 52} (1946) 908.

\bibitem{HPollard48} H. Pollard, {\em The completely monotonic character of the Mittag-Leffler 
function $E_{\alpha}(-x)$}, Bull. Amer. Math. Soc. {\bf 54} (1948) 1115.

\bibitem{IPodlubny99} I. Podlubny, "Fractional Differential Equations", (Academic Press, San Diego, 1999).

\bibitem{TRPrabhakar69} T. R. Prabhakar, {\em A singular integral equation with a generalized 
Mittag-Leffler function in the kernel}, Yokohama Math. J. {\bf 19} (1971) 7--15.

\bibitem{CFAERosa15} C. F. A. E. Rosa and E. Capelas de Oliveira, {\em Relaxation equations: 
fractional models}, J. Phys. Math. {\bf 5} (2015) 1000146.

\bibitem{Sandev19} T. Sandev, \v{Z}. Tomovski, J. L. Dubbeldam, and A. V. Chechkin, {\em Generalized 
diffusion-wave equation with memory kernel}, J. Phys. A: Math. Theor. {\bf 52} (2019) 015201.

\bibitem{AStanislavsky10} {A. Stanislavsky and K. Weron, {\em Anomalous diffusion with under- and overshooting subordination: A competition between the very large jumps in physical and operational times}, Phys. Rev. E {\bf 82} (2010) 051120.}

\bibitem{AStanislavsky15} A. Stanislavsky, K. Weron, and A. Weron, {\em Anomalous diffusion approach 
to non-exponential relaxation in complex physical systems}, Commun. Nonlinear Sci. Numer. Simulat. 
{\bf 24} (2015) 117--126.

\bibitem{AStanislavsky16} {A. Stanislavsky and K. Weron, {\em Atypical Case of the Dielectric Relaxation Responses and its Fractional Kinetic Equation}, Frac. Calc. Appl. Anal. {\bf 19}(1) (2016) 212--228.}

\bibitem{AStanislavsky17} A. Stanislavsky and K. Weron, {\em Stochastic tools hidden behind the empirical dielectric relaxation laws}, Rep. Prog. Phys. {\bf 80} (2017) 036001.

\bibitem{AStanislavsky19} A. Stanislavsky and A. Weron, {\em Control of the transistent subdiffusion 
exponent at short and long times}, Phys. Rev. Research {\bf 1} (2019) 023006.

\bibitem{AStanislavsky19a} A. Stanislavsky and K. Weron, {\em Fractional-calculus tools applied to study 
the nonexponential relaxation in dielectrics} in V. E. Tarasov (ed.) "Handbook of Fractional Calculus 
with Applications. Volume 5. Applications in Physics, Part B" (De Gruyter, Berlin, 2019), 53--70.

\bibitem{AStanislavsky20} A. Stanislavsky and A. Weron, {\em Accelerating and retarding anomalous diffusion: A Bernstein function approach}, Phys. Rev. E {\bf 101} (2020) 052119.

\bibitem{RSchilling10} R. L. Schilling, R. Song, and Z. Vondra\v{c}ek, "Bernstein Functions: Theory 
and Applications" (De Gruyter Studies, Berlin, 2010).

\bibitem{RSchilling} R. L. Schilling, {\em An introduction to L\'evy and Feller processes}, in From 
L\'evy--type processes to parabolic SPDEs. Adv. Courses Math. (CRM Barcelona, Birkh\"auser -- Springer, 
Cham, 2016), 1--126. 

\bibitem{footnote1}{{In mathematical literature the Laplace exponent is better known as the L\'{e}vy, or characteristic exponent of some stochastic process $U$.}}

\bibitem{footnote2} {The role of Bernstein, complete Bernstein and completely monotone functions 
as elucidating problems of anomalous diffusion, in particular as providing tools which allow to judge 
probabilistic interpretation of solutions, has been noticed quite recently and is the subject of 
still growing interest \cite{Sandev19,AStanislavsky20,KGorska20a}.}

\bibitem{footnote2a} {If $\hat{\Psi}(s)$ is CBF for $s\in\mathbb R_>$ then Eq. \eqref{29/05-2} guarantees that the memory {function} $\hat{M}(s)$ is CMF.} 

\bibitem{footnote3}{The inverse Laplace transform of $f(t)$ is given by 
the Bromwich integral $f(t) = L^{-1}[\hat{f}(s); t] = \int_{L} \exp(s t) \hat{f}(s) \D s/(2\pi\!\I)$. 
The (direct) Laplace transform is equal to $\hat{f}(s) = L[f(t); s] = \int_{0}^{\infty} \exp(-t s) 
f(t) \D t$.}

\end{thebibliography}
\end{document}